\documentstyle[preprint,aps,prd,fleqn,epsf]{revtex}

\begin{document}

\draft

\title{Singular Regions in  Black Hole Solutions
       in Higher Order Curvature Gravity}

\author{{S.O. Alexeyev,}\footnote{electronic address: alexeyev@grg2.phys.msu.su}}

\address{Department of Theoretical Physics, Physics Faculty, \\
         Moscow State University, Moscow 119899, RUSSIA}

\author{{M.V. Pomazanov}\footnote{electronic address: michael@math356.phys.msu.su}}

\address{Department of Mathematics, Physics Faculty, \\
         Moscow State University, Moscow 119899, RUSSIA}

\date{June 20, 1997}

\maketitle

\begin{abstract}
     Four-dimensional black hole solutions generated by the low
energy string effective action are investigated outside and
inside the event horizon. A restriction for a minimal black hole
size is obtained in the frame of the model discussed.
Intersections, turning points and other singular points of the
solution are investigated. It is shown that the position and the
behavior of these particular points are definded by various
kinds of zeros of the main system determinant. Some new aspects
of the $r_s$ singularity are discussed.
\end{abstract}

\vskip10mm


\section{Introduction}

     During last years string four-dimensional dilatonic black
holes attracted much attention. As this type of black holes is the
solution of the string theory in its low energy limit, therefore,
by studying these solutions one can hope to clarify some
important unsolved problems of modern theoretical physics, for
example, what is the endpoint of the black hole evaporation, the
quantum coherence and black hole thermodynamics problems
\cite{c1,c2,c3}, so on. After the appearance of the
Gibbons-Maeda-Garfincle-Horowits-Strominger (GM-GHS) solution
\cite{garfin} a great interest to the investigation of the
higher order curvature corrections in the Einstein-dilaton
(Yang-Mills) lagrangian arised
\cite{kanti,maeda,pomaz,stas,donets,mignemi,c15}. It is now very
actual problem because in the regions where the curvature of the
space-time increases the role of the $\alpha'$ expansion
terms grows. The general form of this
expansion is not investigated well yet \cite{c10}. So
as a first step researches study the contribution of the second
order curvature corrections. It was found that the black hole
solutions with the $(\alpha')^1$ terms generalize the wellknown
Schwarzschild solution and the GM-GHS one. The problem is to
find new feathers of the black hole solutions which were
introduced by the string theory.

Some researchers study only the simplified
bosonic part of the low energy string action taken in the form
\begin{eqnarray}\label{e1}
S & = & \frac{1}{16\pi} \int d^4 x \sqrt{-g}
 \biggl[ m^2_{Pl} (-R+2\partial_{\mu} \phi
\partial^\mu \phi ) - \nonumber \\
  & - & e^{-2\phi} F_{\mu\nu} F^{\mu\nu}
+ \lambda \mbox{e}^{-2\phi} S_{GB} \biggr],
\end{eqnarray}
where $R$ is a Ricci scalar; $\phi$ is a dilaton field;
$m_{Pl}$ is the Planck mass;
$F_{\mu\nu}$ is the Maxwell field;
$\lambda=\alpha'/4 g^2$ is the string coupling
parameter describing the Gauss-Bonnet
(GB) term contribution
($S_{GB}=R_{ijkl}R^{ijkl} - 4 R_{ij}R^{ij} + R^2 $)
to the action (\ref{e1}).
It was found that the black hole solution of such action
(with or without the Maxwell term) does exist and provides
the non-trivial dilatonic hair
\cite{kanti,maeda,pomaz,stas,donets,mignemi}.
Further, it was established that the modification of the solution by the
second order curvature corrections became non-vanishing only if the
black hole size (mass) was small enough (the case of the large
value of the coupling constant $\lambda$).
It is necessary to emphasis that the differential equations in
such class models have a very complicated structure. So,
there is no possibility
for the direct analytical solving them.
Hence, the researchers have to use perturbative \cite{mignemi,luosto,natsuume}
or numerical \cite{kanti,maeda,pomaz,stas,donets} methods.

     In our previous papers \cite{pomaz,stas} black hole
type solutions of the action (\ref{e1}) were found by the
special numerical method from infinity up to a some particular
point inside the event horizon $r_h$. It was shown that in the
case when the magnetic charge $q$ was rather small (or vanished)
a singular ``tube'' (in $t$ direction) with the topology $S^2
\times R^1$ has appeared inside the black hole. Asymptotically
flat solution occurs from infinity up to this singular ``tube''
with the radius $r_s$. It takes place in the range of the
magnetic charge $q$ to be $0\leq q < q_{cr}$, where $q_{cr}$ is
a new critical charge value appearing in second order curvature
gravity \cite{stas}. In the case of $q$ to be large enough
($q_{cr} < q < M \sqrt{2}$ in GHS gauge, see \cite{garfin}) the
solution occurs up to zero point where the dilatonic function
$\phi$ diverges as in the GM-GHS case. In this paper we are
going to show that all the particular points (and their main
feathers) of the black hole solution obtained from the set of
implicit ordinary differential equations are defined by the
various kinds of zeros of the main system determinant. In
addition some new feathers of these solutions are discussed.

     The structure of the paper is the following. Sec. II deals
with the main equations, in Sec. III we briefly remind our
previous results and discuss them in the light of the
conclusions of \cite{misha}, in Sec. IV some new physical
feathers of the $r_s$ singularity are presented and Sec. V
contains the main conclusions.

\section{Field Equations}

Our purpose is to study
physical and mathematical feathers of the black hole solutions
(static, spherically symmetric case).
Therefore, the most convenient choice
of metric is
\begin{eqnarray}
ds^2 = \Delta (r) dt^2 - \frac{\sigma^2 (r) }{\Delta (r) } dr^2 - f^2 (r)
(d \theta^2 + \sin^2 \theta d \varphi^2), \nonumber
\end{eqnarray}
where functions $\Delta$, $\sigma$ and $f$
depend only on a radial coordinate $r$.
Various types of the metric gauges were used in our investigations.
The most convenient of them are, so-called, ``curvature gauge''
($f=r$, used in \cite{pomaz}) and GHS gauge ($\sigma = 1$,
used in \cite{stas}).
Corresponding field (Einstein) equations
can be rewritten in the matrix form
\begin{eqnarray}\label{e2}
a_{i1} \Delta '' + a_{i2} f'' + a_{i3} \phi'' & = & b_i,
\qquad \mbox{GHS gauge} \nonumber \\
\tilde{a}_{i1} \Delta '' + \tilde{a}_{i2} \sigma'
+ \tilde{a}_{i3} \phi'' & = & \tilde{b}_i,
\qquad \mbox{``curvature gauge''}
\end{eqnarray}
where $i=1,2,3$ and the coefficients $a_{ij}$ and $b_i$ are equal to
(in the GHS gauge)
\begin{eqnarray}
a_{11} & = & 0, \nonumber \\
a_{12} & = & - m_{Pl}^2 f + 4 \mbox{e}^{-2 \phi} \lambda \phi' \
2 \Delta f', \nonumber \\
a_{13} & = & 4 \mbox{e}^{-2 \phi} \lambda \ (\Delta {f'}^2- 1), \nonumber \\
a_{21} & = & m_{Pl}^2 f + 4 \mbox{e}^{-2 \phi} \lambda \phi' \
2 \Delta f' , \nonumber \\
a_{22} & = & m_{Pl}^2 \  2 \Delta
+ 4 \mbox{e}^{-2 \phi} \lambda \phi' \ 2 \Delta \Delta' ,\nonumber \\
a_{23} & = & 4 \mbox{e}^{-2 \phi} \lambda \ 2 \Delta  \Delta' f' ,
\nonumber \\
a_{31} & = & 4 \mbox{e}^{-2 \phi} \lambda \ (\Delta {f'}^2 - 1) ,
\nonumber \\
a_{32} & = & 4 \mbox{e}^{-2 \phi} \lambda \  2 \Delta \Delta' f' ,
\nonumber \\
a_{33} & = & (-2) m_{Pl}^2 \Delta f^2  ,\nonumber \\
\nonumber \\
b_{1} & = & m_{Pl}^2 f^2 (\phi')^2 \nonumber \\
& + &  4 \mbox{e}^{-2 \phi} \lambda \ (\Delta {f'}^2 - 1) \ 2 (\phi')^2 ,
\nonumber \\
b_{2} & = & (-2) m_{Pl}^2 ( \Delta' f' +  \Delta f (\phi')^2 ) \nonumber \\
& + &
4 \mbox{e}^{-2 \phi} \lambda \ 2 \Delta \Delta'  f' \  2 (\phi')^2
\nonumber \\
& + & 2 \mbox{e}^{-2 \phi} q^2 (1/f^3) \nonumber \\
& - & 4 \mbox{e}^{-2 \phi} \lambda \phi' \ 2 (\Delta')^2 f' ,\nonumber \\
b_{3} & = & 2 m_{Pl}^2 \phi' f \ (\Delta' f + 2 \Delta f') \nonumber \\
& + & 2 \mbox{e}^{-2 \phi} q^2 (1/f^2) \nonumber \\
& - & 4 \mbox{e}^{-2 \phi} \lambda \ (\Delta')^2  {f'}^2 .\nonumber
\end{eqnarray}
The remaining $\delta S / \delta f = 0$ additional equation
fulfills automatically elsewhere on the solution trajectory.

The Arnowitt-Diezer-Misner (ADM) mass $M$ and the
dilaton charge $D$ are defined by the asymptotic
expansions for $\Delta$, $f$ and $\phi$
\begin{eqnarray}\label{e6}
\Delta & = & 1 - \frac{2M}{r} + O\biggl(\frac{1}{r}\biggr); \nonumber \\
f^2 & = & r^2 \Biggl( 1 - \frac{D}{r} +  O\biggl(\frac{1}{r}\biggr)
\Biggr); \\
\phi & = & \phi_\infty + \frac{D}{r} + O\biggl(\frac{1}{r}\biggr). \nonumber
\end{eqnarray}

\section{Investigation of the Particular Points}

     For integrating outside and inside the event horizon, a
method based on integrating over an additional parameter was
used and described in detail in \cite{pomaz}. Here we briefly
remind the main results and discuss their new feathers.

     The results of our numerical integration are depicted in
Fig.1. Shown in the Figure presents the dependence of metric
function $\Delta$ against radial coordinate $r$ with the
different meanings of the event horizon value $r_h$ (in the
curvature gauge). It was found in \cite{stas} that the solution
with the singular turning point $r_s$ exists only if $0 \leq q <
q_{cr}$ where $q_{cr}$ is the critical magnitude of the magnetic
charge. The appearance of $q_{cr}$ is created with the second
order curvature corrections. While $q$ increases and reaches the
$q_{cr}$ value, the solution behavior changes and it looks like
completely to the GM-GHS one. As one can see from Fig.1, the
behavior of the above-mentioned metric function $\Delta$ outside
the regular horizon has the usual form and is similar to the
standard Schwarzschild solution (see, for example,
\cite{kanti,mignemi}). Inside the horizon the solution exists
only down to the value $r=r_s$. Another solution branch
(additional branch) begins from the value $r_s$ and exists only
up to the ``singular'' horizon $r_x$.

     To check the particular points of the system (\ref{e2}) it
is necessary to consider its main determinant (here, for
simplicity, we use ``curvature gauge'' and vanishing value of
magnetic charge; in the GHS gauge with $0 \leq q < q_{cr}$ the
results are completely the same but the formulas have larger
length)
\begin{equation}\label{e12}
D_{main} = \Delta \biggl[ A \Delta^2 + B \Delta + C \biggr] ,
\end{equation}
where
\begin{eqnarray}
A & = & (-32) e^{-4 \phi} \sigma^2 \lambda^2 \biggl[
        4 \sigma^2 {\phi'}^2 m^2_{pl} r^2 - \nonumber \\
        & - & \sigma^2 m^2_{pl}
        + 12 e^{-2 \phi} \Delta' \phi' \lambda \biggr], \nonumber \\
B & = & (-32) e^{-2 \phi} \sigma^4 \lambda \biggl[
        \sigma^2 \phi' m^4_{pl} r^3 + \nonumber \\
        & + & 2 e^{-2 \phi} \sigma^2 \lambda m^2_{pl}
        - 8 e^{-4 \phi} \Delta' \phi' \lambda^2 \biggr], \nonumber \\
C & = & 32 e^{-4 \phi} \sigma^8 \lambda^2 m^2_{pl} -
        2 \sigma^8 m^6_{pl} r^4 + \nonumber \\
        & + & 64 e^{-4 \phi} \sigma^6 \Delta' \lambda^2 m^2_{pl} r
        +128 e^{-6 \phi} \sigma^6 \Delta' \phi' \lambda^3 . \nonumber
\end{eqnarray}

     Eqs. (\ref{e2}) represent the system of ordinary
differential equations in a non-evident form. That is why the
system (\ref{e2}) can have the peculiarities when its main
determinant $D_{main}$ vanishes. The structure of those
peculiarities depends upon the behavior of the eqs. (\ref{e2})
near the singular surface $D_{main}=0$ in phase-space
\cite{misha}. In the model with asymptotically flat solutions
only three types of the $D_{main}$ degeneracy arise. They are
\begin{eqnarray}\label{e13}
&\mbox{(a): }& \Delta = 0, \quad C \neq 0 \qquad\qquad\qquad\qquad\qquad
\mbox{(``intersection''),} \nonumber \\
&\mbox{(b): }& A \Delta^2 + B \Delta + C = 0, \quad
\Delta \neq 0, \quad C \neq 0 \qquad\quad
\mbox{(``turning point''),} \nonumber \\
&\mbox{(c): }& \Delta = 0, \quad C = 0 \qquad\qquad\qquad\qquad\qquad
\mbox{(``complicated singularity''),}
\end{eqnarray}

The case (\ref{e13}a) is realized at the regular horizon $r_h$ (see Fig.1)
and occurs also in the Schwarzschild solution. The asymptotic form of
the solutions near the point $r_h$ has the form \cite{pomaz}
\begin{eqnarray}\label{e3}
\Delta & = & d_1 (r-r_h) + d_2 (r-r_h)^2 + o( (r-r_h)^2), \nonumber \\
\sigma & = & s_0 +s_1 (r-r_h) + o( (r-r_h)), \nonumber \\
\mbox{e}^{-2\phi} & = & \phi_0
(1 - 2 * \phi_1 (r-r_h) + 2 (\phi^2_1 - \phi_2) (r-r_h)^2 ) + o( (r-r_h)^2),
\end{eqnarray}
where $(r-r_h) \ll 1$. Substituting the formulas
(\ref{e3}) into the equations (\ref{e2}), one obtains
the following right relations between the expansion coefficients
($s_0$, $\phi_0$ and $r_h$ are free independent parameters,
here we do correct an unfortunate misprint in our work \cite{pomaz})
\begin{equation}\label{e4}
d_1 ( z_1 d_1^2 + z_2 d_1 + z_3 ) = 0,
\end{equation}
where:
\begin{eqnarray}
z_1 & = & 24 \lambda^2 \phi_0^2 , \nonumber \\
z_2 & = & - m^4_{Pl} r^3_h s_0^2, \nonumber \\
z_3 & = & m^4_{Pl} r^2_h s_0^4, \nonumber
\end{eqnarray}
and the parameter $\phi_1$ for $d_1 \neq 0$ is equal to:
$\phi_1  = \left[ (m_{Pl}^2)/(4 \lambda d_1 \phi_0) \right] *
[r_h d_1 - s_0^2 ]$.

     When $d_1 = 0$, the metric function $\Delta$ has the double
(or higher order) zero. In such a situation the equation for
$d_2$ ($d_3$, $d_4$, $\ldots$) is a linear algebraic one and
there are no asymptotically flat branches.

     When $d_1 \neq 0$ the solution of the black hole type takes
place only if the discriminant of the equation (\ref{e4}) is
greater or equal to zero and, hence, $r_h^2 \geq 2 \lambda \phi_0
\sqrt{14 + 8\sqrt{3}}$. One or two branches occurs and always
one of them is asymptotically flat. With the supposition of
$\phi_\infty =0$ (and, as we tested, in this case
$1.0 \leq \phi_0 < 2.0$)
the {\it infinum} value of the event horizon is
\begin{equation}\label{e9}
r_h^{inf} = \sqrt{\lambda} \ \sqrt{\frac{4 \sqrt{6}}{m^2_{Pl}}} ,
\end{equation}
     The analogous formula in the other interpretation
was studied by Kanti {\it et. al.} in
\cite{kanti}.

The case (\ref{e13}b) is realized inside the regular horizon
at $r=r_s < r_h$ (see Fig.1) as a consequence of the
intensification of the GB-term influence. The solution
behavior strongly differs from the Schwarzschild
one or the GM-GHS one. The $D_{main}$ degeneracy of (\ref{e13}b)
type reduces to the violation
of the uniqueness of the solution at the point $r_s$.
Similar situations are typical for the systems of the type (\ref{e2})
in the neighborhood of the singular surface $D_{main}=0$ \cite{misha}.
The asymptotic behavior of both solution branches near the position $r_s$
can be described by the following formulas using the smooth
function $\sigma$ as an independent variable \cite{pomaz}
\begin{eqnarray}\label{e10}
\Delta & = & d_s + d_2 (\sigma-\sigma_s)^2 + o( (\sigma-\sigma_s)^2) ,
\nonumber \\
r  & = & r_s + r_2 (\sigma-\sigma_s)^2 + o( (\sigma-\sigma_s)^2) ,
\nonumber \\
\exp(-2\phi) & = & \phi_s (1- 2 f_2 (\sigma-\sigma_s)^2 )
+ o( (\sigma-\sigma_s)^2),
\end{eqnarray}
where $\sigma - \sigma_s \ll 1$.
Free independent parameters are: $\sigma_s$, $\phi_s$, $r_s$.
After the substitution of these expansions to the system (\ref{e2}),
one can obtain that $d_2 = f_2$ and the equation for $(d_2/r_2)$
has the form
\begin{equation}\label{e20}
z_4 y^2 + z_5 y + z_6 = 0,
\end{equation}
where $y=(d_2/r_2)$ and the other coefficients are
\begin{eqnarray}
z_4 & = & m^2_{Pl} \sigma_s^2 d_s r_s^2
          + 4 \lambda \phi_s (\sigma_s^2 - 3 d_s), \nonumber \\
z_5 & = & - m^2_{Pl} \sigma_s^2 r_s, \nonumber \\
z_6 & = & m^2_{Pl} \sigma_s^2 (\sigma_s^2 - d_s). \nonumber
\end{eqnarray}
Eq.(\ref{e20}) may have either no solutions
or one or two solutions dependently upon
its discriminant magnitude. The case of the unique solution
corresponds to the minimal value of ${r_h}_{min}$
(eq. (\ref{e9}), see Fig.1 curve (c)).
The situation with the positive discriminant corresponds to the
turning point of the solution. Here it is necessary
to note that if one rewrites the expansions (\ref{e10})
against the expansion parameter $(r-r_s) \ll 1$,
the result will have the following form
$\Delta = d_s + y (r-r_s) + \ldots$ and so on.
Hence, there are only two branches (because of two
possible values of $y$) can exist near the position $r_s$.
They are: the asymptotically flat one
and the ($r_s r_x$) one. {\it No any other solution branches are
in the neighborhood of $r_s$.}
The ``curvature invariant'' $R_{ijkl}R^{ijkl}$ near $r_s$
is equal to
($r - r_s \ll 1$):
$R_{ijkl}R^{ijkl} = T_1 / (r-r_s)
+ o \biggl( 1/ (r-r_s)\biggr) \rightarrow \infty$,
where $T_1=const$.
The components $T^0_0$ and $T_2^2$ of the stress-energy tensor
$T^{\mu}_{\nu}$ also diverges (as $1/\sqrt{r-r_s}$)
near the position $r_s$.

The case (\ref{e13}c) is realized on the singular horizon $r_x$
of the additional branch $(r_s r_x$) (see Fig.1, curves (a) and (b)).
This branch is not asymptotically flat and, hence, is non-physical.
The asymptotic form of this solution near the position $r_x$
is shown in \cite{pomaz}.
``Curvature invariant''
$R_{ijkl}R^{ijkl}$ also diverges near the position $r_x$.

Here it is importantly to stress that the distance between
the points $r_x$ and $r_h$ decreases while decreasing $r_h$ (see Fig. 1).
In the limit point defined by the restriction (\ref{e9})
all particular points pour together $r_h$=$r_s$=$r_x$=${r_h}_{min}$
and the internal structure of the black hole disappears.
The case (\ref{e13}c) is realized in this point and, therefore,
the ``curvature invariant'' diverges. Hence, the point
${r_h}_{min}$ represents the event horizon and the singularity
in the same point. So, such situation contradicts with
the ``cosmic censorship'' hypothesis \cite{torn,c20} but there is a question
about its stability. {\it The possibility of the realization of such situation
is still open.} Returning to the formula (\ref{e9}), one should
remember that $\lambda$ is the combination of the fundamental string
constants. That is why this formula can be reinterpreted as the
restriction to
the minimal black hole size (mass) in the given model.
This restriction appears in the second order
curvature gravity and is absent in the minimal Einstein-Schwarzschild
gravity. This fact can throw an additional light on the problems
of the black holes in our Universe.

\section{$r_s$ singularity}

It is possible to obtain the
approximate relation between $r_s$, $r_h$ and $\lambda$.
Substituting the Schwarzschild values of the metric functions
and the vanishing value of the dilaton charge $D$ to the formula (\ref{e12}),
one obtains
\begin{equation}\label{e14}
r^3_s = \lambda \  \frac{4 \sqrt{3} r_h \phi_s}{m^2_{Pl}}.
\end{equation}
Fig. 2 shows the dependencies of the value $r_s$
against the coupling parameter $\lambda$ given by the formula (\ref{e14})
and by the numerical integration.
From the eq.(\ref{e14}) it is possible to conclude
that the pure Schwarzschild
solution is the limit case of our one with $r_s = 0$.
In the case with rather small value of $\lambda$
this formula gives the good agreement with the results
calculated by the numerical integration. While increasing $\lambda$,
the absolute error increases as a consequence of ignoring
the non-vanishing values $(1 - \sigma)$, $\phi'$ and so on.
It is necessary to point out that the eq.(\ref{e14})
represents the dependence $r_s = $ const $\lambda^{1/3}$
which we suppose to be right
because after the appropriate selection this constant by hands
the agreement between numerical data and this formula improves.
Eq. (\ref{e14}) shows also that when the influence of
the GB term (or black hole mass) increases, $r_s$ also increases.

Further, it is possible to find the approximate relation between
the $q_{cr}$, $\lambda$ and $M$. One can rewrite the $D_{main}$
in the GHS coordinates and substitute there the GM-GHS values
of the metric functions $\Delta$, $f$ and the dilaton function
$\phi$ in the following form
\begin{eqnarray}
\Delta & = & 1 - \frac{2M}{r} , \nonumber \\
f & = & \sqrt{r^2 - r\ \frac{q^2}{M}} , \nonumber \\
e^{-2\phi} & = & 1 - \frac{q^2}{M}\frac{1}{r}.  \nonumber
\end{eqnarray}
Hence, $D_{main}$ takes the form
(we suppose $m^2_{Pl} = 1$ for simplicity)
\begin{eqnarray}\label{ea}
 D_{main} = \frac{T}{ r^{10} M^4 (r M - q^2)^2} ,
\end{eqnarray}
where $T=T(M, \lambda, q^2, r)$ is the polynomial of the sixteenth order
against $r$ (we do not write it because of its dangerous length).
If one supposes the charge $q$ to vanish in the eq. (\ref{ea})
(hence, the denominator of eq. (\ref{ea}) never vanishes) he obtains
the formula (\ref{e14}) in the form $r^3_s = 8 \sqrt{3}
\lambda M$.

The denominator in the eq. (\ref{ea}) can vanish only in the case of
$r = r_k = q^2/M$. If this situation happens, $D_{main}$ diverges.
Consequently, all the senior derivatives
vanishes and then in the point $r_k$ which is also particular one has
only the local minimum (maximum).
So, the condition for the turning point existence
is the following: $r_s$ must be situated righter relatively to
$r_k$, i.e. $r_s \geq r_k$. The  limit condition $r_s = r_k$
is just one for $q_{cr}$ because when $r_k > r_s$
one must have a local minimum (maximum) righter $r_s$.
That is why equating each other the expressions of the $r_s$ and $r_k$
one obtains the approximate formula for $q_{cr}$
\begin{eqnarray}\label{eb}
q_{cr} = \lambda^{1/6} \biggl( 8 \sqrt{3} M^4 \biggr)^{1/6}.
\end{eqnarray}
The last formula analogously to the (\ref{e14}) gives the good
agreement with the data obtained from the numerical
calculations only in the case of rather small $\lambda$.
While increasing $\lambda$ the absolute error increases as well.
This fact can be seen from the Figure 3 showing the dependence
of the value $q_{cr}$ against the coupling constant $\lambda$.
Here
it is necessary to point out once more that the eq.(\ref{eb})
represents the right dependency $q_{cr} = $ const $ \lambda^{1/6}$.

Using the expansions (\ref{e10}), it is possible to study
the behavior of the radial time-like and isotropic geodesics
near the singularity $r_s$ (``curvature gauge'' case with vanishing $q$).
Based on these expansions and on the technics
of the geodesical curve calculations from ref.
\cite{chandr}, after integrating over the radial coordinate
$x=r-r_s \ll 1$ one obtains the expressions for the proper time
$\tau (x)$ and the coordinate time $t (x)$ for the radial
time-like and radial isotropic geodesics
\begin{eqnarray}\label{e16}
\tau (x) & = & \pm C_1 x  + C_2 + \ldots ,
\qquad \mbox{(time-like)} \nonumber \\
t (x)  & = & \pm C_3 x + C_4 + \ldots ,
\qquad \mbox{(time-like)} \nonumber \\
\tau (x) & = & \pm C_5 x + C_6 + \ldots ,
\qquad \mbox{(isotropic)} \nonumber \\
t (x)  & = &  \pm C_7 x + C_8 + \ldots ,
\qquad \mbox{(isotropic)}
\end{eqnarray}
where $C_i$ are the constant values.

Comparing the values $t(x)$ and $\tau(x)$ with the
Schwarzschild ones near the regular event horizon
(see, for example, \cite{chandr}),
one can conclude that the behavior of the
radial geodesics
near the $r_s$ singularity differs from the
radial geodesics behavior of the Schwarzschild solution
near the regular horizon. In our model all the above-mentioned
values limited,
in the Schwarzschild model $t(x)$ everywhere diverges.
The divergence of the $R_{ijkl}R^{ijkl}$ also differs
in the models considered: $1/x^6$ in the Schwarzschild case
near the origin and limited at the horizon
and
only $1/x$ in our case ($x=r-r_s$).
Therefore,
one can suppose that the $r_s$-singularity
is a week one (according to the Clarke classification \cite{clarke})
and it can be removed by the appropriate extension of the metric.
According to the Propositions 8.2.2 and 8.2.3 from \cite{clarke},
this procedure is not forbidden because near the position $v=r_s$
``criterion''
\begin{eqnarray}
K(v) & = & \int\limits_0^v d v' \int\limits_0^{v'} d v'' R_{00} (v'')
\end{eqnarray}
is limited. So, the question on the possibility of removing the
$r_s$ singularity by the appropriate metric extension
remains open. It is necessary to emphasis that the singular turning point
$r_s$ appears in various kinds of metric parametrizations as we tested.

\section{Conclusions}

     The black hole solutions generated by the bosonic part of
the four-dimensional low energy superstring effective action
with the second order curvature corrections are discussed in
this paper. They are obtained by using the special numerical
method described in \cite{pomaz}. It is demonstrated that all
the particular points of the solution, namely, regular horizon
$r_h$, singular horizon $r_x$ and $r_s$-singularity, are defined
by the various types of zeros of the main determinant
$D_{main}$, namely, ``intersection'', ``complicated
singularuty'' and ``turning point'' correspondingly.

     The restriction for a minimal black hole size (mass) is
obtained in the frame of the model with the vanishing Maxwell
field contribution (Einstein-dilaton-Gauss-Bonnet model). This
minimal black hole size (mass) depends upon the combination of
the string fundamental constants.

     The approximate formulas for the $r_s = r_s (\lambda,r_h)$
at vanishing $q$ and $q_{cr} = q_{cr} (\lambda,M)$ are found.
These formulas have a perturbative nature and shows that the
effects of second order curvature corrections become
non-vanishing at rather small sizes of black holes.

     The behavior of the radial time-like and radial isotropic
geodesics is studied. According to this and other criterions
$r_s$ singularity is not ``strong''. We have no arguments that
the $r_s$-singularity is coordinate one (and can be removed by
the appropriate metric extension), but {\it nothing forbids
this.} This question is still open.

     \acknowledgments
     One of the authors (S.A.) would like to thank Professor
D.V.Gal'tsov for useful discussions on the subject of this work.

\newpage

\begin{figure}
\epsfxsize=10cm
\centerline{\epsffile{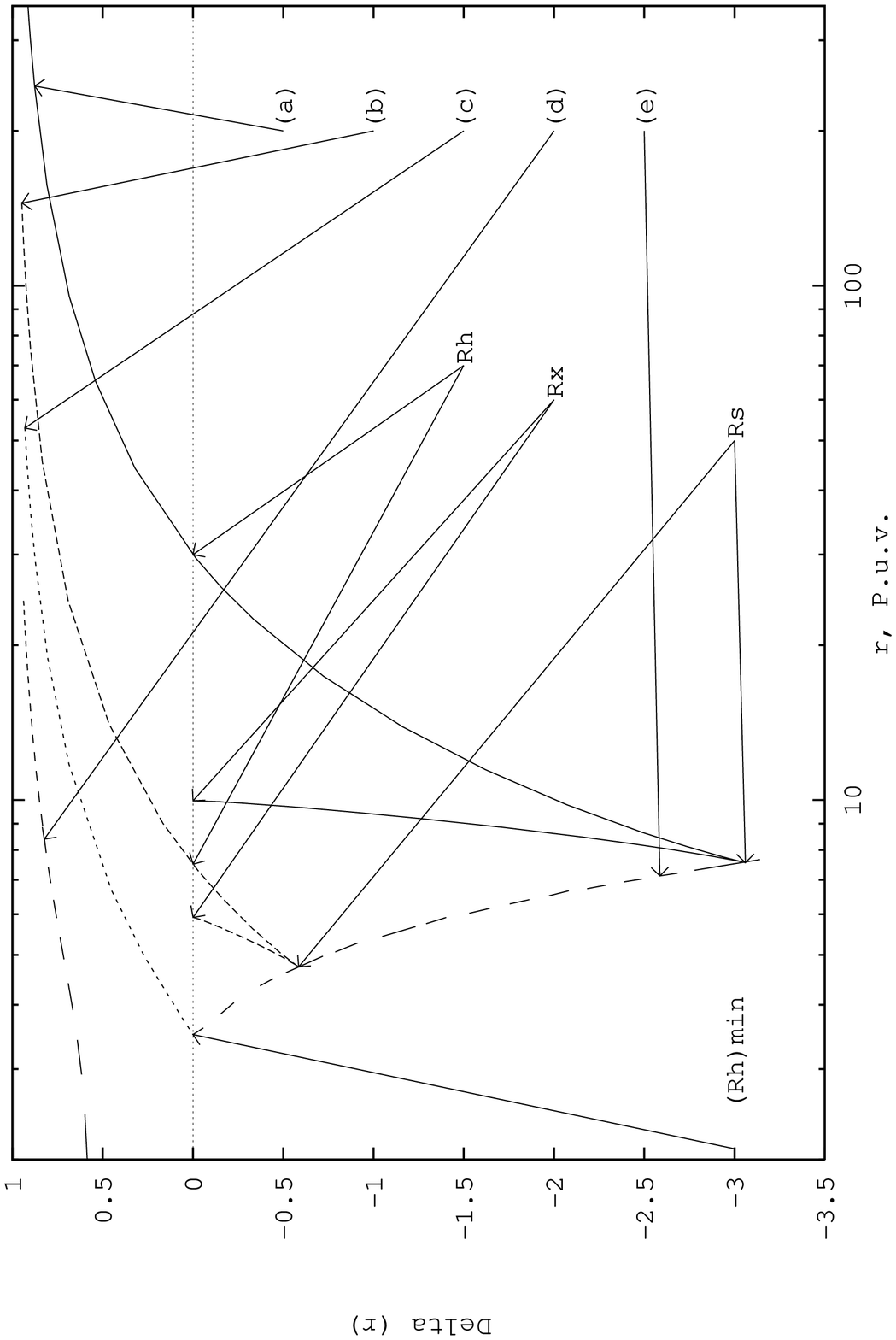}}
\caption{
     The dependence of the metric function $\Delta$ versus the
radial coordinate $r$ at the different values of the event
horizon $r_h$. The ``curvature gauge'' with the vanishing
contribution of the magnetic field is used during the
calculations of the data shown in Figure. The curve (a)
represents the case where $r_h$ is rather large and is equal to
30.0 Plank unit values (P.u.v.). The curve (b) shows the changes
in the behavior of $\Delta(r)$ when $r_h$ is equal to 7.5 P.u.v.
The curve (c) represents the boundary case with
$r_h={r_h}_{min}$ where all the particular points, namely, $r_h$
(regular horizon), $r_s$ (singular turning point) and $r_x$
(singular horizon), pour together and the internal structure
disappears. The curve (d) shows the case where $2M \ll
{r_h}_{min}$ ($2M$=1.5 P.u.v.) and any horizon is absent. The
envelope curve (e) shows the position of the points $r_s$
against the different values of $r_h$.}
\end{figure}

\newpage

\begin{figure}
\epsfxsize=10cm
\centerline{\epsffile{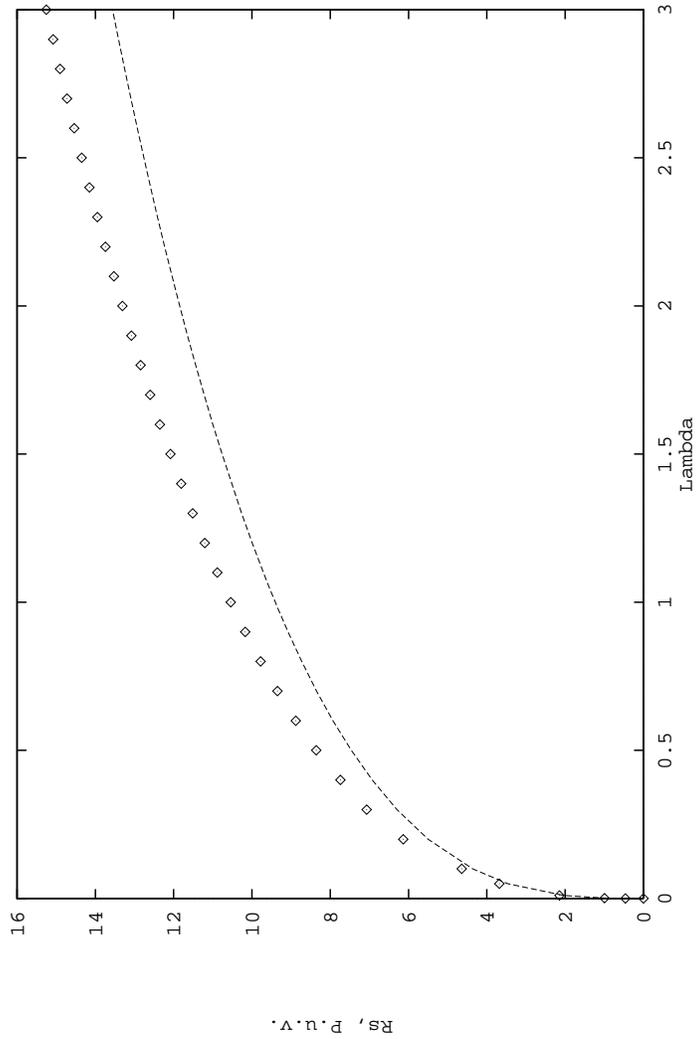}}
\caption{
     The dependence of the position of the singular turning
point $r_s$ versus $\lambda$. Squares represent the values of
$r_s$ calculated from the numerical integration. The solid curve
is obtained by using formula (11) with $r_h=100.0$ P.u.v. and
$\phi_s=1.2$.}
\end{figure}

\newpage

\begin{figure}
\epsfxsize=10cm
\centerline{\epsffile{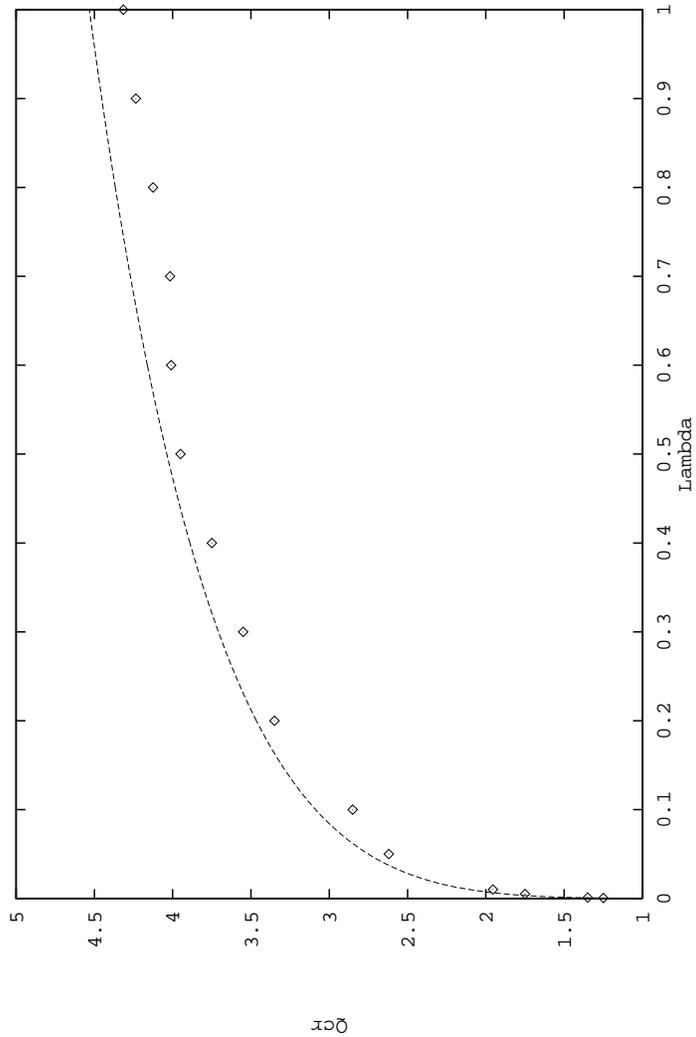}}
\caption{
     The dependence of the position of the critical value of the
magnetic charge $q_{cr}$ versus $\lambda$. Squares represent the
values of $q_{cr}$ calculated from the numerical integration.
The solid curve is obtained by using formula (13) with $M=5.$
P.u.v.}
\end{figure}

\end{document}